\documentclass[sigconf,nonacm]{acmart}

\setcopyright{cc}
\setcctype{by}

\usepackage{setspace}
\usepackage{xcolor}
\usepackage{amsfonts}
\usepackage{amsmath}
\usepackage{amsthm}
\usepackage{enumitem}
\usepackage{multirow}
\usepackage{makecell}
\usepackage{booktabs}
\usepackage{graphics}
\usepackage{algorithm}
\usepackage[noend]{algpseudocode}
\makeatletter
\renewcommand{\theHALG@line}{\thealgorithm.\arabic{ALG@line}}
\makeatother
\usepackage{csquotes}
\usepackage{pifont}
\usepackage{xspace}
\setlist{leftmargin=*}
\usepackage{listings}
\usepackage[noabbrev]{cleveref}
\usepackage{thmtools}
\usepackage{thm-restate}
\usepackage{caption}
\usepackage{subcaption}
\usepackage{libertine}
\usepackage{float}
\usepackage{url}

\newcommand{\codebase}{%
    \url{https://github.com/MystenLabs/walrus}}

\newcommand{\proc}[2]{\textsc{#1}\ensuremath{(#2)}}

\newtheorem{definition}{Definition}
\newtheorem{theorem}{Theorem}

\newtheorem{lemma}{Lemma}

\newcommand{\sysname}{\textsc{Walrus}\xspace}
\newcommand{\redstuff}{\textsc{Red Stuff}\xspace}
\newcommand{\bigO}{\mathcal{O}}
\newcommand{\id}{\ensuremath{\mathit{id}}}
\newcommand{\size}{\ensuremath{\mathit{size}}}
\newcommand{\expiry}{\ensuremath{\mathit{expiry}}}
\newcommand{\abs}[1]{\ensuremath{\lvert #1 \rvert}}
\newcommand{\pri}{\ensuremath{\mathrm{p}}}
\newcommand{\snd}{\ensuremath{\mathrm{s}}}
\DeclareMathOperator{\encode}{Encode}
\DeclareMathOperator{\decode}{Decode}
\newcommand{\await}[1]{\ensuremath{\text{\textbf{await}}_{#1}: \;}}
\newcommand{\awaitfplusone}{\ensuremath{\await{f+1}\{\;}}
\newcommand{\awaittwofplusone}{\ensuremath{\await{2f+1}\{\;}}
\newcommand{\awaitfplusoneindent}
{\hspace{\algorithmicindent}\phantom{\awaitfplusone}}
\newcommand{\awaittwofplusoneindent}
{\hspace{\algorithmicindent}\phantom{\awaittwofplusone}}

\definecolor{pblue}{rgb}{0.13,0.13,1}
\definecolor{pgreen}{rgb}{0,0.5,0}
\definecolor{pred}{rgb}{0.9,0,0}
\definecolor{pgrey}{rgb}{0.46,0.45,0.48}

\definecolor{dullred}{Hsb}{0,1,0.4}
\definecolor{dullyellow}{Hsb}{30,1,0.4}
\definecolor{dullgreen}{Hsb}{60,1,0.4}
\definecolor{dullteal}{Hsb}{150,1,0.4}
\definecolor{dullblue}{Hsb}{210,1,0.4}
\definecolor{dullpurple}{Hsb}{270,1,0.4}
\definecolor{dullmagenta}{Hsb}{300,1,0.4}
\definecolor{dullpurplered}{Hsb}{330,1,0.4}

\definecolor{saffron}{HTML}{F77C00}
\definecolor{ckeyword}{HTML}{7F0055}
\definecolor{ccomment}{HTML}{3F7F5F}
\definecolor{cnumber}{HTML}{2A0099}

\newcommand{\one}{\ding{202}}
\newcommand{\two}{\ding{203}}
\newcommand{\three}{\ding{204}}
\newcommand{\four}{\ding{205}}
\newcommand{\five}{\ding{206}}

\makeatletter
\def\@copyrightpermission{This is the authors' version of a paper published under a CC BY 4.0 license in the Proceedings of the 2026 ACM SIGSAC Conference on Computer and Communications Security (CCS '26). The definitive version is available at \url{https://doi.org/10.1145/3830454.3832580}.}
\makeatother
\begin{document}

\title{\textsc{Walrus}: An Efficient Decentralized Storage Network}

	\author{George Danezis}
	\affiliation{%
		\institution{Mysten Labs \& UCL}
		\city{London}
		\country{UK}
	}
	\email{george@mystenlabs.com}
	\author{Giacomo Giuliari}
	\affiliation{%
		\institution{Mysten Labs}
		\city{London}
		\country{UK}
	}
	\email{giacomo@mystenlabs.com}
	\author{Lefteris Kokoris Kogias}
	\affiliation{%
		\institution{Mysten Labs}
		\city{Athens}
		\country{Greece}
	}
	\email{lefteris@mystenlabs.com}
	\author{Markus Legner}
	\affiliation{%
		\institution{Mysten Labs}
		\city{Zurich}
		\country{Switzerland}
	}
	\email{markus@mystenlabs.com}
	\author{Jean-Pierre Smith}
	\affiliation{%
		\institution{Mysten Labs}
		\city{Zurich}
		\country{Switzerland}
	}
	\email{jean-pierre@mystenlabs.com}
	\author{Alberto Sonnino}
	\affiliation{%
		\institution{Mysten Labs \& UCL}
		\city{London}
		\country{UK}
	}
	\email{alberto@mystenlabs.com}
	\author{Karl Wüst}
	\affiliation{%
		\institution{Mysten Labs}
		\city{Zurich}
		\country{Switzerland}
	}
	\email{karl@mystenlabs.com}

\begin{abstract}
	Decentralized storage faces a fundamental trade-off between replication
overhead, recovery efficiency, and security guarantees. Current approaches
either rely on full replication, incurring substantial storage costs, or
employ erasure-coding schemes that struggle with efficient recovery,
especially under high churn. We present \sysname, a decentralized blob
storage system that addresses these limitations through multiple technical
innovations.

At the core of \sysname is \redstuff, a two-dimensional erasure-coding
protocol that achieves high security with only a $4.5\times$ replication
factor, while providing self-healing of lost data. This means that recovery
is done without centralized coordination and requires bandwidth proportional
to the amount of lost data.

However, \redstuff on its own is not sufficient for \sysname, as it is
designed with a static set of participants in mind. To further support
decentralization, we also introduce a multi-stage epoch-change protocol
that efficiently handles storage node churn while maintaining uninterrupted
availability during committee transitions. Our system incorporates
authenticated data structures to defend against malicious clients and
ensure data consistency throughout storage and retrieval.
\sysname has been deployed in production since March 2025 and has secured 686~TB
of data by July 2026. We conduct an experimental evaluation of the deployed system and demonstrate
that \sysname achieves practical performance at scale and outperforms the Arweave
decentralized storage system.

\end{abstract}


\maketitle
	\renewcommand{\shortauthors}{George Danezis et al.}

\section{Introduction}
Blockchains support decentralized computation through the State Machine Replication (SMR)
paradigm~\cite{schneider1990implementing}. However, they are practically limited to
distributed applications that require little data for operation. Since SMR requires all
validators to replicate data fully, it results in a large replication factor ranging from
$100\times$ to $1000\times$, depending on the number of validators in each blockchain.

While full data replication is needed for computing on state, it introduces
substantial overhead when applications only need to store and retrieve
binary large objects (blobs) without computing on them.\footnote{A recent example includes
	\enquote{inscriptions} on Bitcoin and other chains; see
	\url{https://medium.com/@thevalleylife/crypto-terms-explained-exploring-bitcoin-inscriptions-51699dc218d2}.}
Dedicated
decentralized storage networks~\cite{benisi2020blockchain} emerged to store
blobs more efficiently. For example, early networks like
IPFS~\cite{benet2014ipfs} offer censorship resistance,
reliability, and availability through replication~\cite{zhai2014heading}.

Protocols for decentralized storage generally fall into two main categories. The first
category includes systems using \emph{replication}, with
Filecoin~\cite{psaras2020interplanetary} and Arweave~\cite{williams2019arweave} serving
as prominent examples. The main advantage of these systems is the complete availability
of the blob on selected storage nodes, which allows for easy access and seamless
migration if a storage node goes offline. This setup enables a permissionless environment
since storage nodes do not need to rely on each other for file recovery.

However, the reliability of these systems hinges on the robustness of the selected
storage nodes. For instance, assuming a classic $1/3$ static adversary model and an
unbounded pool of candidate nodes, achieving \enquote{twelve nines} of durability---a
probability below $10^{-12}$ of losing access to a file---requires storing more than 25
copies on the network.\footnote{The chance that all 25 storage nodes are adversarial and
	delete the file is $3^{-25}=1.18 \times 10^{-12}$.}
This results in a $25\times$ storage overhead. A further challenge arises from Sybil
attacks~\cite{douceur2002sybil}, where malicious actors can pretend to store multiple
copies of a file, undermining the system's~integrity.

The second category of decentralized storage services~\cite{li2024sok} uses
\emph{Reed-Solomon (RS) encoding}~\cite{reed1960polynomial}. RS encoding reduces
replication requirements significantly. For example, in a setting similar to that of a
blockchain, with $n$ nodes, of which $1/3$ may be malicious, and in an asynchronous
network, RS encoding can achieve sufficient security with the equivalent of just $3\times$
storage overhead. This is possible since RS encoding splits a file into smaller pieces,
which we call \emph{slivers}, each representing a fraction of the original file. Any set
of slivers greater in total size than the original file can be decoded back into the original file.

However, an issue with erasure coding arises when a storage node fails and needs to be
replaced by another. Unlike fully replicated systems, where data can simply be copied
from one node to another, RS-encoded systems require that all existing storage nodes send
their slivers to the substitute node. The substitute can then recover the lost sliver,
but this process results in $\bigO(\abs{\text{blob}})$ data being transmitted across the
network. Frequent recoveries can erode the storage savings achieved through reduced replication.
As a result, these systems need low churn among storage nodes, which is incompatible
with permissionless settings.

\begin{table*}[t]
	\caption{Comparison of decentralized storage approaches.}
	\label{table:security}
	\centering
	\begin{tabular*}{\textwidth}{@{\extracolsep{\fill}} l c c c c}
		\toprule
		&
		\makecell{\textbf{Replication}\\ \textbf{for $10^{-12}$}} &
		\makecell{\textbf{Write/Read}\\ \textbf{Cost}} &
		\makecell{\textbf{Single Shard}\\ \textbf{Recovery}} &
		\makecell{\textbf{Non-blocking}\\ \textbf{Epoch Change}} \\
		\midrule
		Replication~\cite{psaras2020interplanetary,williams2019arweave}          & $25\times$  & $\bigO(n\abs{\text{blob}})$              &
		$\bigO(\abs{\text{blob}})$               & No  \\
		Classic ECC~\cite{vorick2014sia,storj2018storj}          & $3\times$   & $\bigO(\abs{\text{blob}})$              &
		$\bigO(\abs{\text{blob}})$               & No  \\
		\sysname + \redstuff (\textbf{this work}) & $4.5\times$ & $\bigO(\abs{\text{blob}})$              &
		$\bigO(\abs{\text{blob}}/n)$          & Yes \\
		\bottomrule
	\end{tabular*}
\end{table*}

In this work, we introduce \sysname, a new approach to
decentralized blob storage.
It follows an erasure-coding architecture in order to scale to hundreds of storage nodes, providing
high resilience at a low storage overhead.
At the heart of \sysname lies a new encoding protocol, called \redstuff, which uses a
novel two-dimensional (2D) encoding algorithm that is \emph{self-healing}. Specifically, it enables the recovery of lost slivers using
bandwidth proportional to the amount of lost data ($\bigO(\abs{\text{blob}}/n)$
in our case). Moreover, \redstuff incorporates authenticated data structures to defend
against malicious clients, ensuring that the data remains consistent.
\Cref{table:security} compares \sysname with the two existing approaches.

One additional challenge for \sysname, and in general for any encoding-based decentralized
storage system, is operating securely across epochs,
each managed by a different committee of storage nodes. This is challenging because we
want to ensure uninterrupted availability to both read and write blobs during the
naturally occurring churn of a permissionless system. Yet, if we keep writing data to the
nodes about to depart, the departing nodes must keep transferring that data to the nodes
replacing them. This creates a race for the resources of those nodes, which will either stop
accepting writes or fail to ever transfer responsibility.
\sysname deals with this through its novel multi-stage epoch-change protocol that
naturally fits the principles of decentralized storage systems.

In summary, we make the following contributions:
\begin{itemize}
	\item We define the problem of Asynchronous Complete Data Storage and propose
	      \redstuff, the first protocol to solve it efficiently even under Byzantine faults
	      (\cref{sec:acds}).
	\item We present \sysname, the first permissionless decentralized storage protocol
	      designed for low replication cost and the ability to efficiently recover lost data
	      due to faults or participant churn (\cref{sec:architecture}).

	\item We provide a production-grade implementation of \sysname and deploy both a public
	      testnet and a mainnet.
	      We then measure its performance on testnet and report passive
	      observations from mainnet (\cref{sec:evaluation}).
\end{itemize}

\section{Models and Definitions}\label{sec:model}

\subsection{Network and Adversarial Assumptions}
\sysname runs in epochs, each with a static set of storage nodes.
At the end of the epoch, $n=3f+1$ storage nodes are elected as part of the storage
committee of the next epoch and each one controls a storage \emph{shard} such that a malicious adversary
can control up to $f$ of them.\footnote{For the protocol description, we identify each storage node with a single shard, which lets
us state all thresholds over nodes. In general, the number of shards is a fixed protocol
parameter and a node may control several of them, counting with the corresponding weight
towards every threshold; \cref{sec:reconfiguration,sec:evaluation} describe how shards are
assigned to nodes and reassigned across epochs.} %
The corrupted nodes can deviate arbitrarily from the protocol. The remaining nodes are
honest and adhere to the protocol. If a node controlled by the adversary at
epoch $e$ is not a part of the storage node set at epoch $e+1$, then the adversary can
adapt and compromise a different node at $e+1$ after the epoch change completes.

We assume every pair of honest nodes has access to a reliable and authenticated
channel. The network is asynchronous, so the adversary can arbitrarily delay or reorder
messages between honest nodes, but must
eventually deliver every message unless the epoch ends first. If the epoch ends, then the
messages can be dropped.

\subsection{Erasure Codes}
As part of \sysname, we propose a protocol for Asynchronous Complete Data Storage (ACDS) that
uses a linear erasure-coding scheme. Although it is not necessary for the core parts of the protocol, we also
assume, for some of our optimizations, that the encoding scheme is \emph{systematic},
meaning that the source symbols of the encoding scheme also appear as part of its output symbols.

Let $\encode(B,t,n)$ be the encoding algorithm. Its output is $n$ symbols such that any
$t$ can be used to reconstruct $B$. %
This happens by first splitting $B$ into $t$ symbols of size $\bigO(\abs{B}/t)$
which are called \emph{source} symbols. These are then extended by generating $n-t$
repair symbols for a total of $n$ output symbols. %
On the decoding side, anyone can call $\decode(T, t, n)$ where $T$ is a set of at least
$t$ correctly encoded symbols, and it returns the blob $B$. %

ACDS shares some similarities with Asynchronous Verifiable Information Dispersal
(AVID)~\cite{cachin2005asynchronous,nazirkhanova2022information}, given that the main
goal of both protocols is to distribute data. However, they also have significant
differences; most notably, AVID protocols do not provide completeness, which is critical for
\sysname. A more in-depth discussion is provided in \cref{sec:related}.

\subsection{Blockchain Substrate}
\sysname uses an external blockchain as a black box for all control operations that
happen on \sysname. A blockchain protocol can be abstracted as a computational black box
that accepts a concurrent set of transactions, each with an input message $\mathit{Tx}(M)$ and
outputs a total order of updates to be applied to the state $\mathit{Res}(\mathit{seq},U)$.
We assume that the blockchain does not deviate from this abstraction and does not censor
$\mathit{Tx}(M)$ indefinitely.
Any high-performance modern SMR protocol satisfies these requirements; in our
implementation, we use Sui~\cite{blackshear2024sui}.%

\section{Asynchronous Complete Data Storage (ACDS)} \label{sec:acds}

We first define the problem of (Asynchronous) Complete Data Storage in a distributed system and
describe our solution for an asynchronous network, which we call \redstuff. We then show its
correctness and complexity.

\subsection{Problem Statement}

In a nutshell, a Complete Data Storage protocol allows a writer to write a blob to a
network of storage nodes (\emph{Write Completeness}); ensures that readers read it
consistently despite a potentially malicious writer (\emph{Read Consistency}); and
ensures that any reader can read it despite some failures and malicious behavior among
storage nodes (\emph{Validity}). More formally:

\begin{definition}[Complete Data Storage] \label{def:cds}
	Given a network of $n=3f+1$ nodes,
	where up to $f$ are malicious, let $B$ be a blob that a writer $W$ wants to store
	within the network and share with a set of readers $R$.
	A protocol for Complete Data Storage guarantees three properties:
	\begin{itemize}
		\item Write Completeness: If a writer $W$ is honest, then every honest node holding a
		      commitment to blob $B$ eventually holds a part $p$ (derived from $B$), such that
		      $B$ can be recovered from $\bigO(\abs{B}/\abs{p})$ parts.
		\item Read Consistency: Two honest readers, $R_1$ and $R_2$, reading a successfully
		      written blob $B$ either both succeed and return $B$ or both return $\bot$.
		\item Validity: If an honest writer $W$ successfully writes $B$, then an honest
		      reader $R$ holding a commitment to $B$ can successfully read $B$.
	\end{itemize}
\end{definition}

\begin{definition}[Asynchronous Complete Data Storage]\label{def:acds}
	A Complete Data Storage protocol that is secure in asynchronous networks is an Asynchronous
	Complete Data Storage (ACDS) protocol.
\end{definition}

\subsection{Strawman Design}

In this section, we iterate through two strawman designs and discuss their inefficiencies.

\subsubsection*{Strawman I: Replication}\label{sec:Broadcast-to-all}
The simplest protocol uses replication in the spirit of
Filecoin~\cite{psaras2020interplanetary} and Arweave~\cite{williams2019arweave}. The
writer $W$ broadcasts its blob $B$ along with a binding commitment to $B$ (e.g.,
$H_B = \mathsf{hash}(B)$) to all storage nodes and then waits to receive $f+1$ receipt
acknowledgments.
These acknowledgments form an availability certificate, which guarantees availability as
at least one acknowledgment is from an honest node.
The writer can publish this certificate on the blockchain, which ensures that it is
visible to every other honest node, which can then successfully Read($B$). This achieves
Write Completeness since eventually all honest nodes will hold blob $B$ locally. The rest
of the properties also hold trivially.
Note that the reader never reads $\bot$.

Although the full-replication protocol is simple, it requires the writer to send data of
size $\bigO(n\abs{B})$ over the network, which is also the total cost of storage.
Additionally, if the network is asynchronous, it can cost up to $f+1$ requests to
guarantee a correct replica is contacted, which would lead to $\bigO(n\abs{B})$ cost per
recovering storage node with a total cost of $\bigO(n^2\abs{B})$ over the network. Similarly,
even a read can be very inefficient in asynchrony, as the reader might need to send $f+1$
requests with a cost of $\bigO(n\abs{B})$.

\subsubsection*{Strawman II: Encode \& Share}\label{sec:Reed-Solomon}
\begin{figure}[t]
	\centering
	\includegraphics[width=\columnwidth]{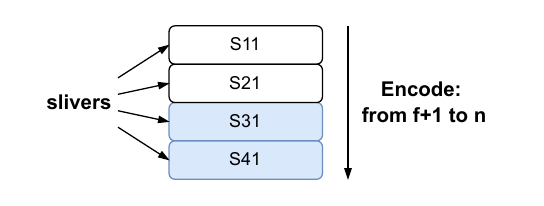}
	\caption{Encoding a blob in one dimension. First, the blob is split into
		$f+1$ systematic slivers and then a further $2f$ repair slivers are encoded.}
	\label{fig:1D}
	\Description{A blob drawn as a horizontal bar divided into $f+1$ systematic slivers,
		followed by $2f$ further slivers produced by erasure encoding, for $3f+1$ slivers in
		total, one per storage node.}
\end{figure}

\looseness=-1
To reduce the upfront data dissemination cost, some distributed storage protocols such as
Storj~\cite{storj2018storj} and Sia~\cite{vorick2014sia} use
RS coding~\cite{reed1960polynomial}. The writer $W$ divides its blob $B$ into $f+1$
slivers and encodes $2f$ extra repair slivers (\cref{fig:1D}). Thanks to the encoding properties, any
$f+1$ slivers can be used to recover $B$. Each sliver has a size of $\bigO(\abs{B}/n)$.
The writer $W$ then commits to the slivers using a binding commitment such as a
Merkle tree~\cite{merkle1988digital} and sends each node a separate sliver together with
a proof of inclusion. %
The nodes receive their slivers and check them against the commitment; if the sliver is
correctly committed, they acknowledge by signing the commitment. The writer $W$
can then generate an availability certificate from $2f+1$ signatures and post it on the blockchain.

A reader continuously requests slivers from the nodes until it receives $f+1$ valid
replies (i.e., replies that are verified against the commitment). The reader is
guaranteed to receive them since at least $f+1$ honest nodes have stored their sliver.
The reader then reconstructs blob $B$ from the slivers and additionally re-encodes
the recovered value and recomputes the commitment~\cite{merkle1988digital,catalano2013vector}.
If writer $W$ was honest, the recomputed commitment will match the commitment from the
availability certificate and the reader outputs $B$. Otherwise, writer $W$ may not have
committed to a valid encoding, in which case the commitments do not match and the reader
outputs $\bot$.

As before, the nodes that did not get slivers during the sharing phase can recover them
by reading $B$. If the output of the read operation is $\bot$, the node returns $\bot$ on
all future reads. Otherwise, the node stores its encoded sliver and discards the rest of $B$.
Note that this recovery process is expensive: recovery costs $\bigO(\abs{B})$ even if the storage
cost afterwards is $\bigO(\abs{B}/n)$.

This second protocol reduces the dissemination costs significantly at the expense of
extra computation (encoding/decoding and committing to slivers from $B$). Disseminating
blob $B$ only costs $\bigO(\abs{B})$%
, which is the same cost as reading it.
However, complete dispersal still costs $\bigO(n\abs{B})$, because, as we saw, the process of
recovering missing slivers requires downloading the entire blob $B$.
Given that there can be up to $f$ storage nodes that did not manage to get their sliver
from writer $W$ and need to invoke the recovery protocol, the protocol has $\bigO(n\abs{B})$ total cost.
This is not only important during the initial dispersal, but also in cases where the
storage node set changes (at epoch boundaries) as the new set of storage nodes needs to
read its slivers by recovering them from the previous set of storage nodes.

\subsection{Final Design: \redstuff} \label{sec:redstuff}
In this section we present the high-level design of \redstuff. The algorithms can be
found in \cref{sec:algorithms} and the proofs in \cref{sec:redstuff-proofs}.

The encoding protocol above achieves the objective of a low overhead factor with very
high assurance, but is still not suitable for a long-lasting deployment.
The main challenge is that in a long-running large-scale system, storage nodes routinely
experience faults, lose their slivers, and have to be replaced. Additionally, in a
permissionless system, there is some natural churn of storage nodes even when they are
well incentivized to participate.

Both of these cases would result in enormous amounts of data being transferred over the
network, equal to the total size of data being stored, in order to recover the slivers for
new storage nodes. This is prohibitively expensive. We instead want the system to
be self-healing such that the cost of recovery is proportional only to the
data that needs to be recovered and scales inversely with $n$.

\redstuff achieves this. It encodes blobs in two dimensions (2D encoding):
The primary dimension is equivalent to the RS encoding used in prior systems. However, in
order to allow efficient recovery of slivers of $B$ we also encode on a secondary
dimension. \redstuff is based on linear erasure coding (see~\cref{sec:model}) and the
Twin-code framework~\cite{rashmi2011enabling}, which provides erasure-coded storage with
efficient recovery in a crash-tolerant setting with trusted writers. We adapt this
framework to make it suitable in the Byzantine fault-tolerant (BFT) setting with a single
set of storage nodes, and we add additional optimizations that we describe further below.

\subsubsection*{Encoding}
\begin{figure}[t]
	\centering
	\begin{subfigure}[t]{\columnwidth}
		\centering
		\includegraphics[width=\columnwidth]{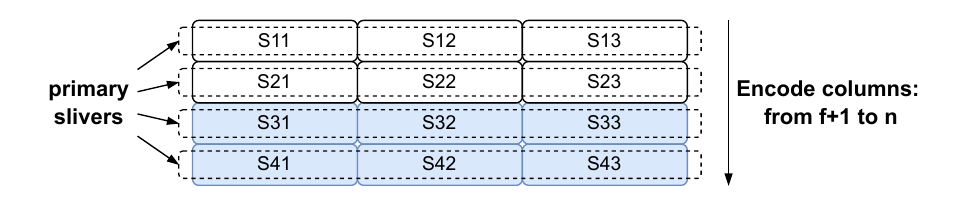}
		\caption{Primary encoding in two dimensions. The blob is split into
			$2f+1$ columns and $f+1$ rows. Each column is encoded as a separate blob with $2f$
			repair symbols. Then each resulting row is the primary sliver of the respective node.}
		\label{fig:2d-primary}
	\end{subfigure}
	\hfill
	\begin{subfigure}[t]{\columnwidth}
		\centering
		\includegraphics[width=\columnwidth]{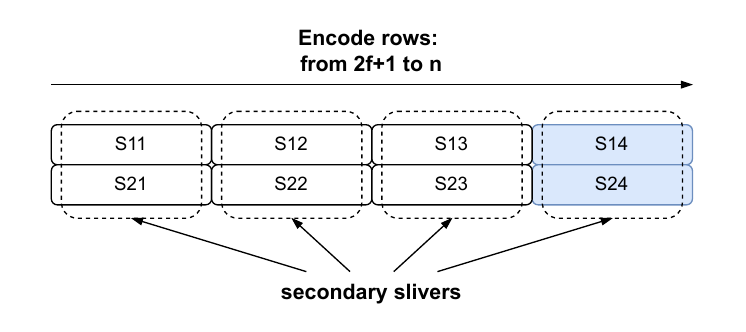}
		\caption{Secondary encoding in two dimensions. The blob is split into
			$2f+1$ columns and $f+1$ rows. Each row is encoded as a separate blob with $f$ repair
			symbols. Then each resulting column is the secondary sliver of the respective node.}
		\label{fig:2d-secondary}
	\end{subfigure}
	\hfill
	\caption{2D encoding / \redstuff.}
	\label{fig:2D}
	\Description{Two matrix diagrams of the same blob. In the first, each column is extended
		with repair symbols and each resulting row forms the primary sliver of a node. In the
		second, each row is extended and each resulting column forms the secondary sliver.}
\end{figure}

Our starting point is the second strawman that splits the blob into $f+1$
slivers. Instead of simply encoding repair slivers, we first add one more dimension to
the splitting process: the original blob is split into $f+1$ primary slivers (horizontal in
the figure) and into $2f+1$ secondary slivers (vertical in the figure). \Cref{fig:2D}
illustrates this process. As a result, the blob is now split into $(f+1)(2f+1)$
symbols that can be visualized in an $[f+1,2f+1]$~matrix. %

Given this matrix we then generate repair symbols in both dimensions. We take each of the
$2f+1$ columns (of size $f+1$) and extend them to $n$ symbols such that there are $n$ rows.
We assign each of the rows as the \emph{primary sliver} of a node
(\cref{fig:2d-primary}). This almost triples the total amount of data we need to send and
is very close to what 1D encoding does in Strawman II.
In order to provide efficient recovery for each sliver, we also take the initial
$[f+1,2f+1]$ matrix and extend each of the $f+1$ rows (of size $2f+1$) with repair symbols
to $n$ symbols (\cref{fig:2d-secondary}) using our encoding scheme. This creates $n$
columns, which we assign as the \emph{secondary sliver} of each node, respectively.

For each sliver, we compute a vector commitment over all symbols of the extended
sliver (i.e., including the repair symbols), forming the \emph{sliver commitment}.
The commitments for every sliver form the blob \emph{metadata}. Using these, nodes can
later, when queried for a single symbol, prove that the symbol they return is the symbol
originally written, which enables Byzantine fault tolerance for \redstuff.
A vector commitment to the metadata then forms the \emph{blob~commitment}.

\subsubsection*{Write Protocol}
The Write protocol of \redstuff uses a pattern similar to that of the RS-code protocol. The writer
$W$ first encodes the blob and creates a sliver pair for each node. A sliver pair $i$ is
the pair of the $i^{\text{th}}$ primary and secondary slivers. There are $n = 3f+1$ sliver
pairs, one per node.

Then, $W$ sends the metadata (i.e., all sliver commitments) to every node, along with the
respective sliver pair. The nodes check their own sliver pair against the commitments,
recompute the blob commitment, and reply with a signed acknowledgment if all checks pass.
Once $2f+1$ signatures are collected, $W$ aggregates them into a certificate and posts it
on-chain to \emph{certify} the blob's availability. The nodes then keep openings of the
blob commitment to their sliver commitments, locally encode the
metadata with a 1D $(f+1)$-out-of-$n$ encoding (along with another vector commitment to the
parts of the encoded metadata), and keep the part assigned to them, together with
the opening of their part.
This is needed since the metadata needs to be available for nodes that need to recover their
slivers (see \emph{Sliver Healing} below). However, fully replicating the metadata would
result in a quadratic overhead, while this approach reduces the overhead to a constant per
node (i.e., from quadratic to linear system-wide overhead).
Our production deployment currently forgoes this optimization and stores the metadata in
full on every node.

In theoretical asynchronous network models with reliable delivery, the above would result
in all correct nodes eventually receiving a sliver pair from an honest writer. However,
in practical protocols, the writer needs to stop retransmitting. It is safe to stop the
retransmission after $2f+1$ signatures are collected, leading to at least $f+1$ correct
nodes (out of the $2f+1$ that responded) holding a sliver pair for the blob.

\begin{figure*}[t]
	\centering
	\begin{subfigure}[t]{0.48\textwidth}
		\centering
		\includegraphics[width=\textwidth,
			trim={65.88bp} {557.28bp} {190.44bp} {180.36bp}, clip]{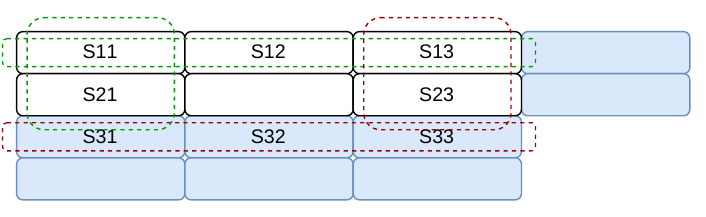}
		\caption{Nodes 1 (green) and 3 (red) hold a primary and secondary sliver each, corresponding to an unextended row and column, respectively.}
		\label{fig:example-1}
	\end{subfigure}
	\hfill
	\begin{subfigure}[t]{0.48\textwidth}
		\centering
		\includegraphics[width=\textwidth,
			trim={65.88bp} {446.22bp} {190.44bp} {291.42bp}, clip]{figures/recovery}
		\caption{Nodes 1 and 3 extend their slivers and send the intersection of the resulting rows/columns with the
			columns/rows of node 4 to node 4 (bold green and red).}
		\label{fig:example-2}
	\end{subfigure}

	\begin{subfigure}[t]{0.48\textwidth}
		\centering
		\includegraphics[width=\textwidth,
			trim={65.88bp} {338.40bp} {190.44bp} {399.24bp}, clip]{figures/recovery}
		\caption{Node 4 uses the $f+1$ symbols on its column to recover the
			missing symbols of its extended secondary sliver (purple). It then sends any other recovering node the
			recovered intersections of this column with their row.}
		\label{fig:example-3}
	\end{subfigure}
	\hfill
	\begin{subfigure}[t]{0.48\textwidth}
		\centering
		\includegraphics[width=\textwidth,
			trim={65.88bp} {230.76bp} {190.44bp} {506.88bp}, clip]{figures/recovery}
		\caption{Node 4 uses the $f+1$ symbols on its row that it received from nodes 1 and 3, symbol S44, which it recovered in the previous step (purple), and
			at least $f-1$ further symbols sent by other nodes after they recover their own secondary slivers (not needed in this example) to recover the missing symbols of its
			primary sliver (teal).}
		\label{fig:example-4}
	\end{subfigure}
	\hfill
	\caption{Nodes 1 and 3 help node 4 recover its sliver pair.}
	\label{fig:recovery}
	\Description{Four panels of the encoding matrix showing node 4 recovering its lost
		sliver pair. Nodes 1 and 3 supply the symbols where their own rows and columns
		intersect node 4's column and row; node 4 first rebuilds its secondary sliver from
		$f+1$ such symbols, then its primary sliver once $2f+1$ symbols are available.}
\end{figure*}

\subsubsection*{Read Protocol}
The Read protocol is similar to the one for one-dimensional RS codes.
It starts with $R$ collecting the metadata, i.e., the list of sliver
commitments for the blob commitment. To do so, $R$ requests the 1D encoded metadata parts
from the nodes along with the opening proofs for the encoded metadata.
Since at least $f+1$ of any $2f+1$ replies originate from honest nodes, the vector
commitment on which the majority of the replies agree is the correct one; parts whose
openings do not verify against it are discarded before decoding.

Since honest nodes only store metadata that they verified against $\id$ when it
	was written, this binds the decoded metadata to the blob.
Then $R$ requests a read for the blob commitment from all nodes and they respond with the
primary sliver they hold (this may happen gradually to save bandwidth). Each response
is checked against the corresponding commitments in the commitment set for the blob. When
$f+1$ correct primary slivers are collected, $R$ decodes $B$, re-encodes it, and
	recomputes the metadata. If it matches the metadata retrieved earlier, then $R$ outputs $B$; otherwise, it outputs~$\bot$.

\subsubsection*{Sliver Healing} The big advantage of \redstuff compared to the
RS-code protocol is its self-healing property. This comes into play when nodes that did
not receive their slivers directly from $W$ try to recover them. Any storage node can
recover its secondary sliver by asking $f+1$ storage nodes for the symbols that exist
in their row, which should also exist in the (extended) column of the requesting node
(\cref{fig:example-1,fig:example-2,fig:example-3}). This means that eventually all $2f+1$
honest nodes will have secondary slivers. At that point, any node can also recover its
primary sliver by asking the $2f+1$ honest nodes for the symbols in their
column~(\cref{fig:example-4}) that should also exist in the (extended) row of the
requesting storage node. In each case, the responding node also sends the opening for the
requested symbol of the commitment of the source sliver. This allows the receiving node
to verify that it received the symbol intended by the writer $W$, which ensures correct
decoding if $W$ was honest.

Since the size of a symbol is $\bigO(\abs{B}/n^2)$, and each storage node will
download $\bigO(n)$ total symbols, the cost per node remains at $\bigO(\abs{B}/n)$
and the total cost to recover the blob is $\bigO(\abs{B})$, which is equivalent to the cost of
a Read and of a Write. As a result, by using \redstuff, the communication complexity of
the protocol is (almost\footnote{Depends on the vector commitment scheme used.})
independent of $n$, making the protocol scalable.

\subsubsection*{Security}
\Cref{sec:redstuff-proofs} provides proofs that \redstuff satisfies all properties
of an ACDS. Informally, Write Completeness is ensured by the fact that a correct writer
will confirm that at least $f+1$ correct nodes received sliver pairs before stopping
retransmissions. The sliver recovery algorithm can then ensure that the remaining honest
nodes can efficiently recover their slivers from these, until all honest nodes eventually
hold their respective sliver, or can prove that the encoding was incorrect. Validity
holds due to the fact that $2f+1$ correct nodes will eventually hold correct sliver
pairs, and therefore a reader that contacts all nodes will eventually get enough slivers
to recover the blob. Read Consistency holds since two correct readers that decode a blob
from potentially different sets of slivers, re-encode it and check the correctness of the
encoding. Either both output the same blob if it was correctly encoded or both output
$\bot$ if it was incorrectly encoded.

\subsubsection*{Discussion on Replication Cost}
The benefits of \redstuff do not come for free. There is an increase in the total storage
cost of the system compared to classic BFT erasure codes, from $3\times$ to $4.5\times$. But this
is necessary, as we need the extra information to allow for efficient recovery. Without
it, the recovery cost for a single sliver would be $\bigO(\abs{B})$ instead of $\bigO(\abs{B}/n)$.

A naive application of our 2D encoding in which nodes store fully extended slivers would
lead to a storage amplification of $9\times$. However, nodes only store a sufficient
number of symbols ($2f+1$ for primary, $f+1$ for secondary) that allows them to locally
decode the rest of the extended sliver when they need to communicate symbols that are not
stored to support another node's recovery, resulting in a factor $4.5\times$ instead.
This trades CPU for storage, which is a favorable trade-off for \sysname as it has
minimal CPU requirements. Note additionally that we do not use a symmetric $[f+1,f+1]$
matrix, as it would result in double the replication cost. Instead, our design achieves
the minimum replication cost at the expense of a slightly more structured recovery procedure.

\section{\sysname: A Decentralized Blob Store}\label{sec:architecture}

The previous section presented \redstuff, an encoding algorithm designed
for efficient recovery. However, an encoding
algorithm alone is not sufficient for a complete storage system. In this
section, we present \sysname, which integrates an encoding algorithm with
a blockchain that serves as a control plane for metadata and governance.
Within an epoch, clients interact with \sysname through two operations:
writing blobs and reading blobs.

\sysname requires an Asynchronous Complete Data Storage (ACDS) protocol and a
blockchain to coordinate storage nodes and track metadata. Our
implementation uses \redstuff (\cref{sec:redstuff}) instantiated with
Reed-Solomon codes~\cite{reed1960polynomial} for the erasure coding, Merkle
trees~\cite{merkle1988digital} for the vector commitments, and the Sui
blockchain~\cite{blackshear2024sui} as the control plane. These choices are not
fundamental; \sysname can use any components that satisfy the requirements in
\cref{sec:model}.

Beyond basic read and write operations, \sysname must also handle an
additional concern: storage nodes may join or leave the system
over time. \Cref{sec:reconfiguration} presents our epoch-change
algorithm, which enables permissionless participation of hundreds of
nodes with seamless handover between committees.

\begin{figure}[t]
	\centering
	\includegraphics[width=\columnwidth]{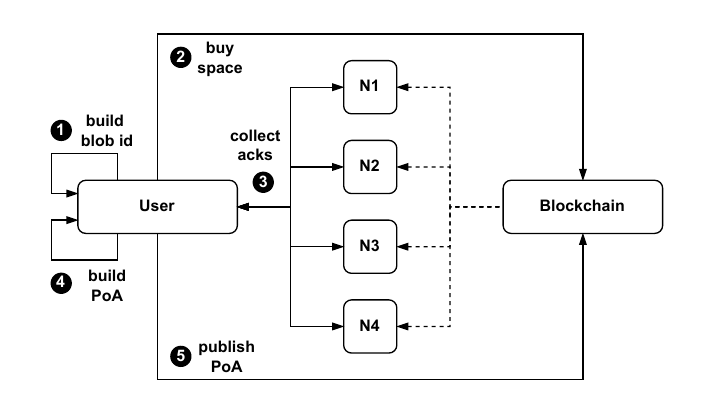}
	\caption{
		\sysname write flow. The writer generates the blob ID of the file
		to store, acquires storage space through the blockchain, submits the encoded file
		to \sysname, collects $2f+1$ acknowledgments, and submits them as proof of
		availability to the blockchain.
	}
	\label{fig:write}
	\Description{A flow diagram of the write path, numbered one to five: the writer encodes
		the blob and derives its blob ID, registers it on the blockchain, sends slivers to the
		storage nodes, collects $2f+1$ signed acknowledgments into a certificate, and publishes
		that certificate on-chain to reach the Point of Availability.}
\end{figure}

\subsection{Writing a Blob}

The process of writing a blob in \sysname can be seen in \cref{alg:store-upd} and
\cref{fig:write}. The process begins with the writer~(\one) encoding a blob using
\redstuff. This process yields sliver pairs, a list of sliver commitments, and a blob
commitment. The writer derives a \emph{blob ID} $\id$ by hashing the blob commitment
with additional metadata such as the length of the file and the type of the encoding
(\cref{alg:helpers-upd}).

Then, the writer~(\two) submits a transaction on the blockchain to acquire sufficient
space for the blob to be stored during a sequence of epochs, and to \emph{register} the
blob. The blob ID and the blob size are sent.
The blockchain smart contract needs to secure sufficient space to store both
the encoded slivers on each node and all metadata associated with the
commitments for the blob. Some payment may be sent along with the transaction to secure
empty space, or empty space over epochs can be a resource that is attached to this
request to be used. Our implementation allows for both options.

Once the register transaction commits~(\three), the writer informs the storage nodes of
their obligation to store the slivers of the blob identified by $\id$, sending them
the transaction together with the commitments and the primary and secondary slivers
assigned to the respective storage nodes along with proofs that the slivers are
consistent with the published $\id$.
This proof is the list of sliver commitments itself: the storage node recomputes $\id$
from the commitments and checks that it matches the registered blob ID; it then recomputes
the commitment of each of the two slivers it received and compares the results against the corresponding entries
(\cref{alg:helpers-upd}).
After the sliver pairs and commitments are verified and stored, the node
responds with a signed acknowledgment over $\id$.

Finally, the writer waits to collect $2f+1$ signed acknowledgments~(\four),
which constitute an availability certificate. This certificate is then published on-chain~(\five),
which denotes the \emph{Point of Availability} (PoA) for the blob in \sysname. The PoA
signals the obligation for storage nodes to keep the slivers available for reads. At
this point, the writer can delete the blob from local storage and go offline.
Additionally, this PoA can be used as proof of availability of the blob by the writer
to third-party users and smart contracts.

Nodes listen to the blockchain for events indicating that a blob reached its PoA. If they
do not hold sliver pairs for this blob, they execute the recovery process to get
commitments and sliver pairs for all blobs past their PoA. This ensures that eventually
all correct nodes will hold sliver pairs for all blobs.

\subsection{Reading a Blob} \label{sec:read}

Reading a blob proceeds in three phases: metadata retrieval, sliver
collection, and reconstruction with verification. Given a blob ID
$\id$, the client first requests the metadata from the
storage nodes. Once it has retrieved the metadata, the client proceeds to request primary slivers.
The client collects primary slivers from storage nodes, verifying each
against the sliver commitments from the metadata. After obtaining $f+1$ valid slivers,
the client reconstructs the blob using erasure decoding.

The final verification step ensures consistency: the client re-encodes
the reconstructed blob and recomputes the metadata. If this matches the
metadata retrieved earlier, the client outputs the blob; otherwise, the encoding
was inconsistent and the client outputs $\bot$. This check guarantees
that all honest readers either obtain the same blob or reject it
entirely---a property that follows directly from \redstuff's guarantees.
In the fault-free case, the total download is only slightly larger than
the original blob size, since primary slivers contain the blob data
with modest redundancy overhead. Moreover, in a steady state, reads are
still available with up to $2f$ faulty shards, as only $f+1$ slivers are
needed.

\subsection{Recovery of Slivers}\label{sec:recovery}

One issue with writing blobs in asynchronous networks or when nodes crash-recover is
that not every node gets their sliver during the write. This is not a problem, as
protocols can function without completeness. Nevertheless,
in \sysname we opted to use a two-dimensional encoding scheme because it allows for
completeness, i.e., the ability of every honest storage node to recover and eventually
hold a sliver for every blob past PoA.
This allows (1) better load balancing of read requests as all nodes can reply to reads and
(2) dynamic availability of nodes, which enables reconfiguration without needing
to reconstruct and rewrite every blob.

All these benefits rely on the ability of storage nodes to recover their slivers
efficiently. The protocol closely follows the \redstuff recovery protocols illustrated in
\cref{fig:recovery}. When a storage node sees a certificate of a blob for which
it has not received slivers, it tries to recover its sliver pair from the rest of the
storage nodes. For this, it requests from all storage nodes the symbols corresponding to
the intersection of the recovering node's primary/secondary sliver with the signatory
nodes' secondary/primary slivers.
Given that $2f+1$ nodes signed the certificate, at least $f+1$ will be honest and reply.
This is sufficient for all $2f+1$ honest nodes to eventually hold their secondary
slivers. As a result, when all honest nodes hold their secondary slivers, they can share
those symbols corresponding to the recovering nodes' primary slivers; those nodes then
reach the $2f+1$ threshold and also recover their primary slivers.

\subsection{Handling Byzantine Faults during Encoding}\label{sec:inconsistency}
One last challenge for \sysname is handling Byzantine faults, both for writers and for
storage nodes. First, storage nodes could misbehave by acknowledging slivers
that they do not hold or by serving wrong slivers during read/recovery. We defend against
these attacks through the commitment to slivers that the writer creates, meaning that any
invalid sliver will be rejected (\cref{alg:helpers-upd}, line~\ref{alg:line:verify_sliver}).

Second, a malicious client could upload slivers that are inconsistent, i.e., do not correspond
to a correct encoding of a blob. In that case, a node may not be able to recover a sliver that
is consistent with the commitment of the symbols that it received.
\sysname does not require certified blobs to be recoverable (which is the primary goal of
AVID~\cite{cachin2005asynchronous}), as the writer can always encode recoverable garbage
data. Instead, it is guaranteed that unrecoverable blobs will have a
third-party-verifiable proof of inconsistency, associated with $\id$, after a read fails.

The read process executed by a correct reader rejects any inconsistently encoded blob by
default (\cref{alg:helpers-upd}, line~\ref{alg:line:decode}), and as a result, sharing this
proof is not a necessity to ensure consistent reads. However, agreeing on the
inconsistency allows nodes to delete this blob's data and thus save space and bandwidth.

To prove inconsistency, the storage node shares the
inconsistency proof---consisting of the symbols that it received for recovery and their
inclusion proofs---with the other nodes, who can verify it by performing a trial recovery
themselves. After verifying this fraud proof, each verifying node attests on-chain that $\id$ is
invalid. After observing $f+1$ such attestations, all nodes will subsequently
respond with $\bot$ to any request for the inconsistent blob's slivers, along with a
pointer to the on-chain evidence for the inconsistency. %

This means that even if a malicious writer colludes with $f$ malicious nodes to
certify an invalid blob, the rest of the storage nodes will detect this during a read or
recovery and produce an inconsistency proof, guaranteeing agreement among honest readers.

\subsection{Committee Reconfiguration} \label{sec:reconfiguration}

\looseness=-1
The number of shards is a protocol parameter, fixed at 1000 in the current deployment, and
$n=3f+1$ counts shards rather than operators. The number of shards therefore does not grow with
the number of nodes in the committee in practice: shards are assigned to nodes in proportion to
their delegated stake, so a larger committee only changes how the same shards are distributed and
never requires re-encoding.
The security invariant is stated over shards as well, requiring that the adversary control at most
$1/3$ of them. \sysname is orthogonal to the mechanism that enforces or incentivizes this
invariant: the current deployment uses delegated proof of stake, but proof of work, proof of
personhood, or random committee sampling from a larger population would serve equally well.

\sysname is a decentralized protocol; therefore, it is natural that the set of storage
nodes will fluctuate between epochs.
When a new committee replaces the current committee between epochs, reconfiguration takes place.
The goal of the reconfiguration protocol is to preserve the invariant that all blobs past
the PoA are available, regardless of whether the set of storage nodes
changes. Additionally, \sysname must continue to perform reads and writes for blobs to
ensure no downtime even if the reconfiguration process takes hours.\footnote{Unlike
	blockchain protocols where reconfiguration simply transfers responsibility for validation,
	in \sysname it transfers responsibility for data storage on the order of terabytes or even petabytes.}

\subsubsection*{Core Design}
\looseness=-1
At a high level, the reconfiguration protocol of \sysname is similar to the
reconfiguration of blockchain systems, since \sysname also operates in quorums of storage nodes.
However, the reconfiguration of \sysname has its own challenges because the migration of
state is orders of magnitude more expensive than in classic blockchain systems. The most
important challenge is the race between writing blobs for epoch $e$ and transferring
slivers from outgoing storage nodes to incoming storage nodes during the reconfiguration
event between $e$ and $e+1$. More specifically, if the amount of data written in epoch
$e$ exceeds what the departing storage node can transfer to
the incoming storage node, then the epoch will never finish. This problem is exacerbated
when some of the outgoing storage nodes of $e$ are unavailable, as this means that the
incoming storage nodes need to recover the slivers from the~committee.

To resolve this problem without shutting off the write path, we take a different approach
than blockchain protocols. Specifically, we require writes to be directed to the
committee of $e+1$ the moment the reconfiguration starts, while still directing reads to
the old committee, instead of having a single point at which both reads and writes are
handed over to the new committee. We can afford to do this, unlike blockchains, because
\sysname does not handle its state consistency (which is the responsibility of Sui in our
deployment) and hence can have divergence and duplication.
However, this creates challenges when it comes to reading fresh blobs, as during the
handover period it is unclear which nodes store the data. We handle this by including in
the metadata of every blob the epoch in which it was first written. If the epoch is
$e+1$, then the client is asked to direct reads to the new committee; otherwise, it can
direct reads to the old committee. This happens only during the handover period (when
both committees need to be live and~secure).

Once a member of the new committee has bootstrapped its part of the state, i.e., it has
received all slivers for its shards, it signals that it is ready to take over.
When nodes of the new committee collectively holding $2f+1$ shards have signaled this,
the reconfiguration process terminates and all reads are redirected to the storage nodes
of the new committee.

\subsubsection*{The Importance of \redstuff for Reconfiguration}
As we mentioned above, a significant challenge arises during epoch change when some nodes are
faulty. In prior work, a single faulty node would require bandwidth equal to the size of
the file to be transferred over the network of storage nodes. This makes epoch changes
prohibitively expensive and is the reason no prior decentralized system has one.
The key enabler of \sysname's ability to handle this gracefully is our \redstuff algorithm, as it
allows for the bandwidth cost of the faulty case to be on the same order as that of the
fault-free case.
We discuss the associated overheads in \cref{sec:evaluation}.

\subsubsection*{Security Arguments} In a nutshell, reconfiguration ensures all ACDS properties
across epochs. The key invariant is that the reconfiguration algorithm ensures that if a blob
is to be available across epochs, in each epoch $f+1$ correct shards (potentially
different ones) hold slivers. This is the purpose of the explicit signaling, by nodes collectively
holding $2f+1$ shards, that unlocks the epoch change. Therefore, eventually all
other honest storage nodes can recover their sliver pairs, and in all cases, honest
nodes in the next epoch are able to recover correct sliver pairs for $f+1$ shards as a
condition to move epochs.

\section{Detailed Algorithms} \label{sec:algorithms}
\begin{algorithm}[tb]
	\caption{\sysname helper functions.}
	\label{alg:helpers-upd}
	\small

	\begin{algorithmic}[1]
		\State \texttt{nodes} \Comment{committee of storage nodes, $\abs{\texttt{nodes}}=n=3f+1$}

		\Statex
		\Procedure{EncodeBlob}{$B$}
		\State $E \gets \Call{SplitIntoMatrix}{B, (f+1) \times (2f+1)}$ \Comment{the source symbols}
		\State $E \gets [\Call{ErasureEncode}{E_{(*,j)}, f+1, n}: j \in [0,2f+1)]^\top$
		\Statex \Comment{extend columns: $[n \times (2f+1)]$}
		\State $E \gets [\Call{ErasureEncode}{E_{(i,*)}, 2f+1, n}: i \in [0,n)]$
		\Statex \Comment{extend rows: $[n \times n]$}
		\State $S^\pri \gets [E_{(i,*)}: i \in [0,n)]$ \Comment{primary slivers, one per node}
		\State $S^\snd \gets [E_{(*,i)}: i \in [0,n)]^\top$ \Comment{secondary slivers, one per node}
		\State \Return $(S^\pri,S^\snd)$
		\EndProcedure

		\Statex
		\Procedure{MakeMetadata}{$S^\pri,S^\snd$}
		\State $M^\pri \gets [\Call{MerkleTree}{S^{(\pri,i)}}: i \in [0,n)]$
		\State $M^\snd \gets [\Call{MerkleTree}{S^{(\snd,i)}}: i \in [0,n)]$
		\State $M \gets (M^\pri,M^\snd)$
		\State \Return $M$
		\EndProcedure

		\Statex
		\Procedure{MakeBlobId}{$M, \size$}
		\State $(M^\pri,M^\snd) \gets M$
		\State $(h^\pri, h^\snd) \gets (\Call{MerkleTree}{M^\pri}, \Call{MerkleTree}{M^\snd})$
		\State $h \gets \Call{Hash}{h^\pri, h^\snd}$ \Comment{the blob commitment}
		\State $\id \gets \Call{Hash}{h, \size}$
		\State \Return $\id$
		\EndProcedure

		\Statex
		\Procedure{EncodeMetadata}{$M$}
		\State $m \gets \Call{ErasureEncode}{M, f+1, n}$ \Comment{$n$ metadata shares}
		\State $\tau \gets \Call{MerkleTree}{m}$
		\State \Return $[(i, m_i, \tau, \Call{Opening}{m, i}): i \in [0,n)]$ \Comment{one piece per node}
		\EndProcedure

		\Statex
		\Procedure{DecodeMetadata}{$P$} \Comment{metadata pieces $(i, m, \tau, \pi)$}
		\State $\tau^\star \gets \Call{Majority}{[\tau : (i,m,\tau,\pi) \in P]}$ \Comment{by honest majority}
		\State $V \gets \{(i,m) : (i,m,\tau,\pi) \in P,\ \Call{VerifyOpening}{\tau^\star, m, \pi, i}\}$
		\Statex \Comment{verified shares}
		\State \Return $\Call{ErasureDecode}{V, f+1, n}$
		\EndProcedure

		\Statex
		\Procedure{VerifySliver}{$S^{(x,i)}, M$} \Comment{sliver type $x \in \{\pri,\snd\}$}
		\label{alg:line:verify_sliver}
		\State $(M^\pri,M^\snd) \gets M$
		\State \Return $\Call{MerkleTree}{S^{(x,i)}} = M^x_i$
		\EndProcedure

		\Statex
		\Procedure{DecodeBlobFromPrimary}{$\{S^{(\pri,*)}\}_{f+1}, M$}
		\State $B \gets \Call{ErasureDecode}{\{S^{(\pri,*)}\}_{f+1}, f+1, n}$
		\State $(S^\pri, S^\snd) \gets \Call{EncodeBlob}{B}$ \Comment{re-encode}
		\State $M' \gets \Call{MakeMetadata}{S^\pri,S^\snd}$
		\If{$M' \neq M$} \Return $\bot$ \EndIf \label{alg:line:decode}
		\State \Return $B$
		\EndProcedure
	\end{algorithmic}
\end{algorithm}

This section supplements \cref{sec:architecture} by providing detailed algorithms for
client and storage node operations (\cref{alg:client-upd,alg:store-upd}).
\begin{algorithm}[tb]
  \caption{\sysname client operations.}
  \label{alg:client-upd}
  \small

  \begin{algorithmic}[1]
    \State \texttt{nodes}

    \Statex
    \Statex // Store a blob
    \Procedure{StoreBlob}{$B, \expiry$}
    \State $(S^\pri,S^\snd) \gets \Call{EncodeBlob}{B}$
    \State $M \gets \Call{MakeMetadata}{S^\pri,S^\snd}$
    \State $\size \gets \Call{ByteSize}{B}$
    \State $\id \gets \Call{MakeBlobId}{M, \size}$
    \State $\Call{ReserveBlob}{\id, \size, \expiry}$
    \State $R \gets \{\;\}$
    \For{$i \in [0,\abs{\texttt{nodes}})$}
    \State $N \gets \texttt{nodes}[i]$
    \State $\textsf{StoreRqst} \gets (\id, \size, M, S^{(\pri,i)}, S^{(\snd,i)})$
    \State $R \gets R \cup \{(N, \textsf{StoreRqst})\}$
    \EndFor
    \State $C_A \gets \await{2f+1}\{ c \gets \Call{Send}{N, r}: (N,r) \in R \}$ \Comment{signed acks}
    \State $\Call{StoreCertificate}{C_A, \id}$ \Comment{the availability certificate}
    \EndProcedure

    \Statex
    \Statex // Retrieve metadata
    \Procedure{RetrieveMetadata}{$\id$}
    \State $\textsf{MetadataRqst} \gets (\id)$
    \State $P_M \gets \await{2f+1}\{ p \gets \Call{Send}{N, \textsf{MetadataRqst}}:$
    \Statex \awaittwofplusoneindent\phantom{$P_M \gets {}$}%
    $N \in \texttt{nodes} \}$ \Comment{metadata pieces}
    \State \Return $\Call{DecodeMetadata}{P_M}$
    \EndProcedure

    \Statex
    \Statex // Read from primary %
    \Procedure{ReadBlob}{$\id$}
    \State $M \gets \Call{RetrieveMetadata}{\id}$
    \State $\textsf{SliversRqst} \gets (\id)$
    \State $\{S^{(\pri,*)}\}_{f+1} \gets \awaitfplusone S^{(\pri,i)} \gets \Call{Send}{N, \textsf{SliversRqst}}:$
    \Statex \awaitfplusoneindent $N=\texttt{nodes}[i],\; \Call{VerifySliver}{S^{(\pri,i)}, M} \;\}$
    \State $B \gets \Call{DecodeBlobFromPrimary}{\{S^{(\pri,*)}\}_{f+1}, M}$
    \State \Return $B$
    \EndProcedure
  \end{algorithmic}
\end{algorithm}

In addition to the helper functions specified in \cref{alg:helpers-upd},
these algorithms also leverage the following (intuitive) functions:
\proc{Byte\-Size}{B} to compute the size of a blob $B$ in bytes;
\proc{Erasure\-Decode}{T, t, n} to erasure-decode the source data from a set $T$ of at
least $t$ out of $n$ correctly encoded symbols or slivers (\cref{sec:model});
\proc{Erasure\-Encode}{D, t, n} to erasure-encode data $D$ into $n$ symbols such that
any $t$ of them can recover $D$ (\cref{sec:model});
\proc{Hash}{\cdot} to compute a cryptographic hash;
\proc{Inconsistency\-Proof}{T} to assemble from the collected symbols and openings $T$
a third-party-verifiable proof that the encoding is inconsistent (\cref{sec:inconsistency});
\proc{Majority}{v} to return the most frequent element of a vector $v$;
\proc{Merkle\-Tree}{v} to compute a Merkle tree over a vector input $v$;
\proc{Node\-Index}{N,\texttt{nodes}} to obtain the index of node $N$ in the committee;
\proc{Opening}{v, i} to compute the opening for the entry at index $i$ of the vector
commitment to $v$;
\proc{Request\-Symbol}{m,\id,(r,c)} to request the symbol at position $(r,c)$ of the blob
$\id$ from node $m$, together with an opening against the commitment of the sliver
that contains it;
\proc{Send}{N, \mathit{msg}} to send the message $\mathit{msg}$
to node $N$ over the authenticated channel;
\proc{Sign}{x} to sign $x$ with the node's key;
\proc{Split\-Into\-Matrix}{D, r \times c} to reshape data $D$ into a matrix of the
specified size; and
\proc{Verify\-Opening}{\tau, x, \pi, i} to verify with the opening $\pi$ that $x$ is the
entry at index $i$ of the vector committed to by $\tau$.

Furthermore, the client and storage nodes use the following functions to interact with
the blockchain:
\proc{Is\-Registered}{\id} to check if a blob ID $\id$ is registered on the blockchain;
\proc{Read\-Certificate}{\id} to read an availability certificate for blob ID $\id$ from
the blockchain;
\proc{Reserve\-Blob}{\id, \size, \expiry} to acquire storage for a blob of size $\size$
until epoch $\expiry$ and register the blob ID $\id$ on the blockchain; and
\proc{Store\-Certificate}{C_A, \id} to store the availability certificate $C_A$ for blob
ID $\id$ on the blockchain.
\begin{algorithm}[tbp]
  \caption{\sysname store operations.}
  \label{alg:store-upd}
  \small

  \begin{algorithmic}[1]
    \State \texttt{N} \Comment{identifier of the storage node}
    \State $\texttt{db}_m$ \Comment{persists this node's encoded metadata part}
    \State $\texttt{db}_b$ \Comment{persists slivers}

    \Statex
    \Statex // Store slivers
    \Procedure{StoreSlivers}{$\textsf{StoreRqst}$}
      \State $(\id, \size, M, S^{(\pri,i)}, S^{(\snd,i)}) \gets \textsf{StoreRqst}$
      \Comment{$i$ is this node's index}
      \If{$\neg \Call{IsRegistered}{\id}$} \Return $\bot$ \EndIf
      \State $\id' \gets \Call{MakeBlobId}{M, \size}$
      \If{$\id' \neq \id$} \Return $\bot$ \EndIf
      \If{$\neg \Call{VerifySliver}{S^{(\pri,i)}, M}$} \Return $\bot$ \EndIf
      \If{$\neg \Call{VerifySliver}{S^{(\snd,i)}, M}$} \Return $\bot$ \EndIf
      \State $(M^\pri, M^\snd) \gets M$
      \State $\texttt{db}_m[\id] \gets (\Call{EncodeMetadata}{M}_i,$
      \Statex \hspace{\algorithmicindent}\phantom{$\texttt{db}_m[\id] \gets ($}%
      $\Call{Opening}{M^\pri, i}, \Call{Opening}{M^\snd, i})$
      \Statex \Comment{metadata piece and openings of the blob commitment}
      \State $\texttt{db}_b[\id] \gets (S^{(\pri,i)}, S^{(\snd,i)})$ \Comment{may be truncated (\cref{sec:redstuff})}
      \State \Return $\Call{Sign}{\id}$ \Comment{signed acknowledgment}
    \EndProcedure

    \Statex
    \Statex // Serve slivers (primary path for reads)
    \Procedure{ServeSlivers}{$\textsf{SliversRqst}$}
      \State $\id \gets \textsf{SliversRqst}$
      \If{$\neg \Call{ReadCertificate}{\id}$}
        \State \Return $\bot$ \Comment{ensure on-chain availability certificate exists}
      \EndIf
      \State $(S^{(\pri,i)}, S^{(\snd,i)}) \gets \texttt{db}_b[\id]$
      \State \Return $S^{(\pri,i)}$ \Comment{serve primary sliver for read threshold $f+1$}
    \EndProcedure

    \Statex
    \Statex // Recover one sliver of this node from verified intersections
    \Procedure{RecoverSliver}{$\id, \texttt{nodes}, i, M, x$} \Comment{sliver type $x \in \{\pri,\snd\}$}
      \Statex \hspace{\algorithmicindent}// ask each node $m$ with index $j$ for the symbol shared by
      \Statex \hspace{\algorithmicindent}// $S^{(x,i)}$ and $S^{(\bar x,j)}$, opened against $M^{\bar x}_j$
      \State $Q \gets \await{t_x}\{ (j,z,\pi) \gets \Call{RequestSymbol}{m, \id, p_x(i,j)} :$
      \Statex \hspace{\algorithmicindent}\phantom{$Q \gets \await{t_x}\{$}$m \in \texttt{nodes},\; j = \Call{NodeIndex}{m,\texttt{nodes}},$
      \Statex \hspace{\algorithmicindent}\phantom{$Q \gets \await{t_x}\{$}$\Call{VerifyOpening}{M^{\bar{x}}_j, z, \pi, i} \;\}$
      \State $X \gets \{(j,z) : (j,z,\pi) \in Q\}$ \Comment{verified symbols}
      \State $S_0^{(x,i)} \gets \Call{ErasureDecode}{X, t_x, n}$ \Comment{source symbols}
      \State $S^{(x,i)} \gets \Call{ErasureEncode}{S_0^{(x,i)}, t_x, n}$ \Comment{fully extended sliver}
      \If{$\Call{MerkleTree}{S^{(x,i)}} \neq M^x_i$}
        \State \Return $(\bot, \Call{InconsistencyProof}{Q})$ \Comment{see \cref{sec:inconsistency}}
      \EndIf
      \State \Return $(S^{(x,i)}, \bot)$
    \EndProcedure

    \Statex
    \Statex // Recover slivers for this node (symbol-level)
    \Procedure{RecoverSlivers}{$\id, \texttt{nodes}$}
      \If{$\neg \Call{ReadCertificate}{\id}$} \Return $\bot$ \Comment{only past the PoA} \EndIf
      \State $M \gets \Call{RetrieveMetadata}{\id}$ \\\Comment{required to verify openings; defined in \cref{alg:client-upd}}
      \State $(M^\pri, M^\snd) \gets M$
      \State $i \gets \Call{NodeIndex}{\texttt{N}, \texttt{nodes}}$ \Comment{this node's row/column index}
      \State $(S^{(\snd,i)}, \varphi) \gets \Call{RecoverSliver}{\id, \texttt{nodes}, i, M, \snd}$ \Comment{secondary first}
      \If{$\varphi \neq \bot$} \Return $\varphi$ \Comment{inconsistency proof} \EndIf
      \State $(S^{(\pri,i)}, \varphi) \gets \Call{RecoverSliver}{\id, \texttt{nodes}, i, M, \pri}$
      \If{$\varphi \neq \bot$} \Return $\varphi$ \Comment{inconsistency proof} \EndIf

      \State $\texttt{db}_b[\id] \gets (S^{(\pri,i)}, S^{(\snd,i)})$
      \Comment{persist both slivers}
      \State $\texttt{db}_m[\id] \gets (\Call{EncodeMetadata}{M}_i,$
      \Statex \hspace{\algorithmicindent}\phantom{$\texttt{db}_m[\id] \gets ($}%
      $\Call{Opening}{M^\pri, i}, \Call{Opening}{M^\snd, i})$
      \State \Return $\textsf{ok}$
    \EndProcedure
  \end{algorithmic}
\end{algorithm}

\Cref{tab:notations} summarizes the main notations used in the algorithms. Subscripts of
matrices and vectors denote access to a specific index.

\begin{table}[tb]
  \centering
  \caption{Main notations used in the \sysname algorithms.}
  \begin{tabular}{rl}
    \toprule
    $E_{(i,j)}$           & Symbol at position $(i,j)$ of an encoded blob   \\
    $S^\pri$                 & The set of primary slivers                      \\
    $S^\snd$                 & The set of secondary slivers                    \\
    $S^{(\pri,i)}$           & The primary sliver held by storage node $i$     \\
    $S^{(\snd,i)}$           & The secondary sliver held by storage node $i$   \\
    $\{S^{(\pri,*)}\}_{f+1}$ & Any set of $f+1$ primary slivers                \\
    $S_0^{(x,i)}$            & The source symbols of the sliver $S^{(x,i)}$    \\
    $\bar{x}$               & The sliver type other than $x$, i.e., $\bar\pri = \snd$ \\
    $t_x$                   & Reconstruction threshold: $t_\pri = 2f+1$, $t_\snd = f+1$ \\
    $p_x(i,j)$              & Position of the symbol shared by $S^{(x,i)}$ and $S^{(\bar x,j)}$ \\
    $M$                      & The blob metadata $(M^\pri, M^\snd)$            \\
    $M^\pri$                 & Metadata associated with the primary slivers    \\
    $M^\snd$                 & Metadata associated with the secondary slivers \\
    $h$                      & The blob commitment over the metadata $M$       \\
    $\id$                    & The blob ID, i.e., \proc{Hash}{h, \size}         \\
    \bottomrule
  \end{tabular}
  \label{tab:notations}
\end{table}

\section{\redstuff Proofs} \label{sec:redstuff-proofs} %
This section formally proves that \redstuff satisfies all properties of an ACDS scheme. %

\subsection{Write Completeness}
We show that \redstuff satisfies Write Completeness. Informally, if an honest writer
writes a blob $B$ to the network, every honest storage node eventually holds a correctly
encoded primary and secondary sliver of $B$. For this part we assume the writer is
honest and provides a correct vector commitment $M$.

\begin{lemma}[Primary Sliver Reconstruction] \label{lm:primary-reconstruction}
	If a party holds a set of $(2f+1)$ symbols $\{E_{(i,*)}\}_{2f+1}$ from a primary sliver
	$S^{(\pri,i)}$, it can obtain the complete primary sliver $S^{(\pri,i)}$.
\end{lemma}
\begin{proof}
	The proof directly follows from the reconstruction property of erasure codes with
	reconstruction threshold $(2f+1)$.
\end{proof}

\begin{lemma}[Secondary Sliver Reconstruction] \label{lm:secondary-reconstruction}
	If a party holds a set of $(f+1)$ symbols $\{E_{(*,i)}\}_{f+1}$ from a secondary sliver
	$S^{(\snd,i)}$, it can obtain the complete secondary sliver $S^{(\snd,i)}$.
\end{lemma}
\begin{proof}
	The proof directly follows from the reconstruction property of erasure codes with
	reconstruction threshold $(f+1)$.
\end{proof}

\begin{theorem} \label{thm:write-completeness}
	\redstuff satisfies Write Completeness (\cref{def:cds}).
\end{theorem}
\begin{proof}
	To write a blob $B$, an honest writer $W$ sends at least $(2f+1)$ correctly encoded
	slivers (parts) to different storage nodes, along with a binding vector commitment $M$
	over those slivers. For these nodes the property holds by definition. Now let us assume
	a node $i$ that is not in the initial $2f+1$ recipients. The node will ask every node
	$j$ for the shared symbols in its primary (i.e., $E_{(i,j)}$) and secondary (i.e.,
	$E_{(j,i)}$) sliver. Given the binding vector commitment $M$, node $j$ can either send the
	true symbols or not reply. Since at least $2f+1$ nodes acknowledged $M$, node $i$
	will get $f+1$ correct symbols for its primary sliver $\{E_{(i,*)}\}_{f+1}$ and $f+1$
	correct symbols for its secondary sliver $\{E_{(*,i)}\}_{f+1}$. From
	\cref{lm:secondary-reconstruction}, this means that $i$ will reconstruct its full
	secondary sliver $S^{(\snd,i)}$.

	Since this reasoning applies to any generic node $i$, it holds for all nodes. As a
	result, eventually all $2f+1$ honest nodes will reconstruct their secondary slivers
	$S^{(\snd,*)}$. Every time a node $j$ reconstructs its secondary sliver, it also replies to
	node $i$ with the shared symbol, which is part of the primary sliver of $i$ (i.e.,
	$E_{(i,j)}$). As a result, eventually $i$ will go from $\{E_{(i,*)}\}_{f+1}$ to $\{E_{(i,*)}\}_{2f+1}$.
	This allows node $i$ to apply \cref{lm:primary-reconstruction} and reconstruct its
	primary sliver $S^{(\pri,i)}$.

	Since this reasoning applies to any node $i$, it holds for all nodes and
	concludes the proof that all honest nodes will eventually hold both their primary and
	secondary slivers.
\end{proof}

\subsection{Read Consistency}
We prove that \redstuff satisfies Read Consistency. Informally, if two honest readers
read a blob $B$ stored on the network, they either both eventually obtain $B$ or both
eventually fail and obtain $\bot$.

\begin{theorem}
	\redstuff satisfies Read Consistency (\cref{def:cds}).
\end{theorem}
\begin{proof}
	Note that the encoding scheme is deterministic and the last step of reading is to
	re-run the encoding and reconstruct $M$. As a result, a reader that accepts the read as
	correct needs to output $B$.

	The challenge with Read Consistency is whether the writer can convince two readers
	that collect different slivers to output $B$ and $\bot$, respectively. Let us assume that two honest
	readers $R_1$ and $R_2$ read a blob $B$ from the network and $R_1$ eventually obtains
	$B$ while $R_2$ eventually obtains~$\bot$.

	There are two scenarios for $R_2$ to output $\bot$:
	\begin{enumerate}
		\item $R_2$ gets $f+1$ replies that match the vector commitment $M$ and tries to reconstruct.
		      After reconstruction, re-encoding, and recomputing the vector commitment, it does
		      not match $M$.
		\item Some node failed to reconstruct its secondary sliver. By the algorithm, this
		      node will hold a proof of inconsistency, which it will send to $R_2$. %

	\end{enumerate}

	In either scenario, $R_1$ should also have detected the inconsistency during the
	reconstruction and output $\bot$; otherwise, the binding property of the vector
	commitment would not hold. Hence a contradiction.
\end{proof}

\subsection{Validity}
We prove that \redstuff satisfies Validity. Informally, if an honest writer writes a
correctly encoded blob $B$ to the network, every honest reader eventually obtains $B$.

\begin{theorem}
	\redstuff satisfies Validity (\cref{def:cds}).
\end{theorem}
\begin{proof}
	To write a blob $B$, an honest writer $W$ constructs $n$ correctly encoded slivers (parts)
	along with a binding vector commitment $M$ over those slivers. Since the writer is
	honest, from \cref{thm:write-completeness}, at least $2f+1$ honest storage nodes
	will hold their respective slivers. Let $\texttt{nodes}$ denote the entire set of
	storage nodes.
	An honest reader queries each storage node $N \in \texttt{nodes}$ for its primary
	slivers, verifies each against $M$, and, once it holds $f+1$, uses them to reconstruct
	$B$. Since all honest storage nodes will eventually reply to the reader and $W$ was
	honest, the reader will eventually obtain~$B$.
\end{proof}

\section{Evaluation} \label{sec:evaluation}

Our evaluation aims at demonstrating the following claims:
\begin{enumerate}[label=\textbf{C\arabic*}]
	\item \label{claim:latency} \sysname achieves \emph{low latency}, bounded by network delay.
	\item \label{claim:throughput} \sysname clients achieve \emph{high read and write throughput}.
	\item \label{claim:reconfiguration} \sysname \emph{reconfigures and recovers} without downtime.

\end{enumerate}

To make the evaluation as realistic as possible, we perform measurements on the public \sysname deployments described below, thus exposing the system to real-world conditions, real users, and infrastructure outside our control.

\subsection{Production Deployments}

The \sysname protocol has powered a decentralized production testnet since October 2024 (redeployed in April 2025), and a mainnet since March 2025.
The storage nodes and clients on these networks are written in asynchronous multi-core Rust. They use axum~\cite{axum} for HTTPS networking, fastcrypto~\cite{fastcrypto} for cryptography, rocksdb~\cite{rocksdb} for storage, reed-solomon-simd~\cite{reed-solomon-simd} for erasure coding, and Sui~\cite{sui} as a fast blockchain. The code is open-source and available at \codebase.

Both networks use the same encoding parameters with 1000 shards across approximately 100 storage nodes, although the number of nodes and their shard distribution varied slightly over time. The following values were observed on 2026-07-29:\footnote{Committee, shard, and capacity values are read from the on-chain state. Blob counts, registered and active sizes, and peak-day figures are based on our index of \sysname events on Sui. Country distribution is based on geolocation of the nodes' public IP addresses. Hardware information was self-reported by operators in a questionnaire. Some of these values are also shown on \url{https://walrus.chainviz.io}.}

\subsubsection*{Mainnet}

The \sysname mainnet comprises 95 storage nodes in 17 countries, the top 5 in terms of the total stake being the USA, Finland, the Netherlands, Germany, and Lithuania. No single country has more than $1/3$ of the stake. Besides Europe and North America, nodes are also located in the APAC and LATAM regions. Nodes hold between 4 and 26~shards (median: 7, P90: 22).

Most storage nodes run a recent LTS version of Ubuntu with at least 16~CPU cores, 128~GB RAM, and 1~Gbps bandwidth. Hardware varies across Intel and AMD CPUs and HDD, SSD, and NVMe storage. Nodes offer between 7.0~TB and 2.03~PB of storage (median: 55~TB, P90: 100~TB, P99: 362~TB). The total system capacity (encoded size) is 4.00~PB, 56\% of which is currently used.\footnote{The \enquote{system capacity} here refers to the contract-enforced capacity based on a vote by the committee. The total amount of storage reported by nodes in the committee is 7.86~PB.} Aside from the large storage capacity, nodes are modestly provisioned similarly to typical quorum-based blockchains~\cite{sui-specs}.

Since the public launch in March 2025, 17.6~million unique blobs have been registered by 9,675 wallets with a total unencoded size of 686~TB, 477~TB of which is still active. The maximum number of blobs certified in one day was around 834,000 on 2025-04-02 (approximately 10 blobs per second). The maximum total unencoded size of blobs certified in one day was 21.0~TB on 2026-01-29 (corresponding to an average goodput of approximately 1.95~Gbps).

\subsubsection*{Testnet}

The \sysname testnet comprises 101 storage nodes with a geographical and stake distribution similar to that of the mainnet. Node specifications are slightly lower, with most nodes having at least 8~CPU cores and 64~GB RAM. Storage capacities vary between 322~GB and 75~TB (median: 5.0~TB, P90: 16.5~TB), resulting in a system capacity of 333~TB.

\subsection{Baselines}

The other two main operational decentralized storage networks are Filecoin and Arweave.
To compare \sysname with Filecoin, we focus on two metrics: (1) the time until a client obtains an incentivized cryptographic guarantee that a file has been durably stored, and (2) the time to read the sealed (secured) copy, as opposed to a transient cached \enquote{hot} copy. Because Filecoin relies on resource-intensive proofs of storage~\cite{fisch2018poreps}, the latency to reach these guarantees is (within sector limits, up to 32~GiB) largely independent of file size. \Cref{tab:arweave-and-filecoin} reports these latencies: 1.5~hours to write and seal, and 3~hours to read the sealed copy~\cite{filecoin-latency}.

\begin{table}[t]
	\centering
	\caption{Time to write and read on Filecoin~\cite{filecoin-latency} and to store until finality on Arweave for 10~kB, 10~MB, 20~MB, 70~MB, and 135~MB files. Arweave reads are shown in \cref{fig:latency}.}\label{tab:arweave-and-filecoin}
	\begin{tabular}{llr}
		\toprule
		\textbf{System} & \textbf{Operation}                & \textbf{Latency} \\
		\midrule
		Filecoin        & Write \& secure seal              & 1.5~h            \\
		                & Read from secured copy            & 3~h              \\
		Arweave         & Store to finality (10~kB)         & 38~min           \\
		                & Store to finality (10~MB--135~MB) & 35--36~min       \\
		\bottomrule
	\end{tabular}
\end{table}

Arweave guarantees storage through its mining algorithm, which takes tens of minutes and allows it to quickly fulfill storage promises when compared to Filecoin.
\Cref{tab:arweave-and-filecoin} illustrates the end-to-end latency to store blobs of varying sizes on Arweave. Write latency averages 35 to 38~minutes, regardless of the blob's size, highlighting that the bottleneck is in the mining algorithm.

The following sections demonstrate that \sysname's separation of consensus from data storage results in latencies that are orders of magnitude smaller than those of both Filecoin and Arweave.

\subsection{System Performance} \label{sec:sysname-performance}

We evaluated \sysname's performance on the \emph{public testnet} by collecting latency and throughput measurements as of 2026-01-12 from an AWS \texttt{m5d.8xlarge} instance (10~Gbps bandwidth, 32~vCPUs, 128~GB RAM) located in US East (N. Virginia). This client was used for evaluation and is not part of the \sysname testnet.

\subsubsection*{\sysname Latency}

\begin{figure}[t]
	\centering
	\includegraphics[width=0.96\columnwidth]
	{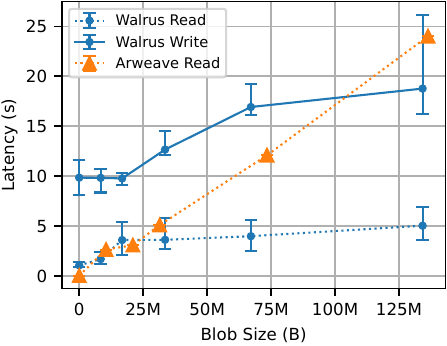}
	\caption{Median read and write latencies for various unencoded blob sizes on \sysname and read latencies on Arweave.  Arweave writes require minutes and so are presented in \cref{tab:arweave-and-filecoin}. Error bars indicate the 5th and 95th percentile latencies.}
	\label{fig:latency}
	\Description{A line plot of median read and write latency against unencoded blob size
		for \sysname, alongside Arweave read latency, with error bars for the 5th and 95th
		percentiles. \sysname reads stay a few seconds across all sizes and writes stay under
		20 seconds, while Arweave reads are substantially slower.}
\end{figure}

\Cref{fig:latency} illustrates the end-to-end latencies of \sysname for 50 stores and reads of random blobs of various sizes. Up to four blobs were written or read concurrently; all blobs were written before any were read. The measurements span from before the client encodes the blob until it creates a PoA on the blockchain, which is after finality but may be before all 1000~shards have received their respective slivers.

The read latencies of both \sysname and Arweave were low, even for large blobs. On \sysname, blobs up to 135~MB were read in less than 5~seconds. By contrast, Arweave's read latencies were around 25~seconds for the 135~MB files. This difference follows from how the two systems retrieve data: a \sysname client requests $f+1$ primary slivers concurrently from several different storage nodes (\cref{alg:client-upd}), each carrying roughly $1/(f+1)$ of the blob, so the download aggregates the bandwidth of the nodes it contacts and is ultimately limited by the client's own link. An Arweave client instead downloads the entire file from a single node, whose resources are also shared with the consensus protocol.

Median write latencies on \sysname were higher than read latencies and remained less than 20~seconds for blobs up to 135~MB. For small blobs, less than 20~MB, the write latencies were consistently around 10~seconds. Both were significantly lower than Arweave's write to finality, which required over 30~minutes (\cref{tab:arweave-and-filecoin}).

\begin{figure}[t]
	\centering
	\includegraphics[width=0.96\columnwidth]
	{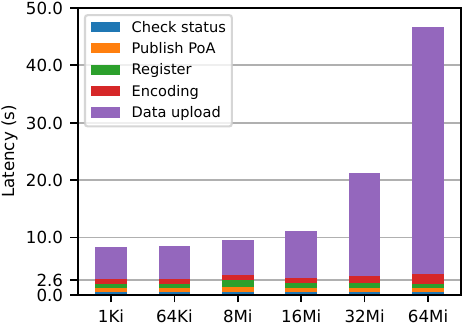}
	\caption{Latencies for various operations during a single blob upload for various sizes. The data upload latency indicates the latency to upload to all online shards, as opposed to the minimum necessary quorum.}
	\label{fig:latency-breakdown}
	\Description{A stacked breakdown of write latency by blob size across the five steps
		of an upload: encoding, check status, register, data upload, and publish PoA. The
		constant-cost steps stay small, and data upload grows with blob size.}
\end{figure}

\Cref{fig:latency-breakdown} provides more details on \sysname's write latencies for the various blob sizes. Each write operation consists of five key steps: \emph{encoding} (erasure encoding the blob), \emph{check status} (checking prior storage status of the blob), \emph{register} (reserve space for the blob), \emph{data upload} (uploading slivers to storage nodes and retrieving the signed acknowledgments that form the availability certificate), and \emph{publish PoA} (committing the PoA to the blockchain).

These operations either have constant latencies: check status, register, publish PoA; or variable latencies: encoding and data upload. For these six blob sizes, the constant latencies were less than 2.6~seconds. In contrast, most of the write time was dominated by the uploads of the slivers and metadata and the retrieval of the acknowledgments. For small blobs, this upload latency was primarily composed of the overhead for the constant-sized metadata uploads, but grew linearly from around 8~MiB upward. Therefore, these results validate our claim~\ref{claim:latency}: \sysname achieves low latency and is bounded by network delays.

\subsubsection*{Client Throughput}

\begin{figure}[t]
	\centering
	\includegraphics[width=0.96\columnwidth]
	{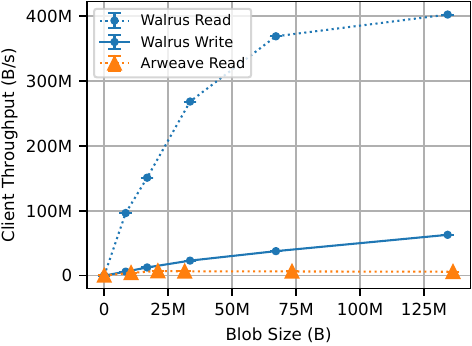}
	\caption{Single-client throughput of \sysname and Arweave for various unencoded blob sizes.}\label{fig:client-tps}
	\Description{A plot of single-client throughput in bytes per second against unencoded
		blob size for \sysname and Arweave. \sysname read throughput rises with blob size to
		around 400~MB/s and write throughput to around 62~MB/s, while Arweave plateaus near
		7~MB/s.}
\end{figure}

\Cref{fig:client-tps} illustrates the throughput (in bytes per second) that a single \sysname and Arweave client can achieve. The read throughput of \sysname was measured by requesting stored blobs at a constant rate for 10--20~seconds (depending on the blob size) before increasing the request rate by 30--60 requests/minute and repeating. The throughput was then the highest constant request rate before the first read failure (e.g., timeout). The write throughput was measured by performing uploads with a constant 8 concurrent requests, the client's configured concurrency, which was taken as a reasonable default for a commodity system. In both cases, the reported throughput is the request throughput multiplied by the blob size. Arweave's write throughput is omitted: its client communicates only with a central gateway, so we could not measure it reliably.

\looseness=-1
As expected, \sysname's read throughput increases with the blob size, as fewer requests for larger amounts of data better utilize the network connections between the client and the storage nodes. Accordingly, the single client achieved around 400~MB/s throughput for 135~MB blobs. By contrast, Arweave's read throughput was severely limited by the blockchain interactions.
The \sysname client's write throughput linearly increased to around 62~MB/s. The limiting factor here was the client's configuration of 8 workers and concurrent requests. This could be relaxed until the bottleneck becomes the user's CPU or network, as each worker maintains its own blockchain state and Sui supports a much higher throughput in transactions per second. However, this configuration is already sufficient to saturate many residential connections. In contrast, Arweave's client plateaued at around 7~MB/s, which is the limit of the consensus protocol to agree on storing data. As such, these results validate claim~\ref{claim:throughput}: \sysname enables clients to read and write at high throughput.

\subsection{Epoch-Change Overhead}

The main overhead associated with the transition from one \sysname epoch to another arises from shards being relocated across epochs, that is, when a shard assigned to a node in epoch $e$ is reassigned to a different node in epoch $e+1$.

The transfer of a shard to another node requires sending slivers of all active blobs and, if the new owner was not part of the committee in epoch $e$, all blob metadata.
During a \emph{shard transfer}, the previous owner of the shard sends the shard's slivers to the new owner.
If the previous owner is unwilling or unable to do so, the receiving node instead performs \emph{shard recovery}.

Shard reassignments on mainnet have been frequent but mostly small. Across the 36 epochs between the launch in March 2025 and July 2026, twelve epoch changes reassigned five or more shards: epochs 2, 6, 9, 11, 16, 20, 23, 27, 28, 30, 32, and 36. The largest reassignment moved 27~shards at epoch 2 (2025-04-08), when the network was young and held 4.4~million blobs; the largest since then moved 13~shards, at epochs 28 (2026-04-07) and 36 (2026-07-28). In at least three cases, shards were moved from a node that was no longer available, requiring the new owners to \emph{recover} their new shards (epochs 17, 20, and 23). We describe two shard \emph{transfers} and two shard \emph{recoveries} in further detail.\footnote{Committee memberships and shard assignments are computed from the system object at each epoch. Sliver counts, transfer volumes, and durations are based on Prometheus metrics voluntarily reported by node operators, which is also how we distinguish transferred from recovered shards.}

At epoch 9 (2025-07-15), a joining node received 4~shards. For the 9.9~million blobs that were live at the time, a total of 636~GB of blob metadata was transferred over 12.5~hours at around 113~Mbps; and 79.5~million slivers, 890~GB, over 15.25~hours at around 129~Mbps.
At epoch 11 (2025-08-12), a node holding 12~shards left the committee and its shards were reassigned to twelve different nodes, one each. Every receiving node was already in the previous committee and therefore held the metadata, so only slivers had to be transferred: 6.6~million primary and 6.6~million secondary slivers per shard, about 210~GB each and 2.5~TB in total. Since each receiver synchronized independently of the others, up to seven shards were transferred concurrently, at an aggregate rate of roughly 1.4~Gbps on average and 2.5~Gbps at peak, and the last shard completed about four hours after the epoch change.

We also analyze two epoch changes that required recovery because the node leaving the committee was no longer online to hand over its slivers.
At epoch 17 (2025-11-04), a node holding 4~shards was removed, and each of its shards held slivers for approximately 465,000 blobs.\footnote{The drop in the number of blobs relative to earlier epochs is due to the fact that users started bundling multiple files within a single blob around June 2025 to reduce overheads and cost, and earlier blobs expired over time.} Three of these shards were reassigned to a single node, which recovered them sequentially in 17 to 21~hours each, and the fourth was recovered by another node.
At epoch 20 (2025-12-16), a node holding 7~shards was removed and its shards were distributed across seven nodes, each recovering slivers for approximately 521,000 blobs. Five shards were recovered concurrently, which took approximately 64~hours in total. Recovery of one of the other shards took 16~hours; the seventh recovery was interrupted and resumed only after its node came back online.

These cases demonstrate that \sysname can efficiently recover even from the complete failure of a node holding several shards. Note that in all of these cases, \sysname remained available for storing and reading data, thus validating claim~\ref{claim:reconfiguration}.

\section{Related Work}\label{sec:related}

\looseness=-1
IPFS~\cite{benet2014ipfs} provides a decentralized store for files, and is extensively
used by blockchain systems and decentralized apps for their storage needs. It
provides content-addressable storage for blocks, and uses a distributed hash table (DHT)
to maintain a link between file replicas and nodes that store them. Publishers must
pin files to storage nodes, usually for a fee, to ensure the files remain available. The underlying storage uses replication on a few nodes for each file.

Filecoin~\cite{psaras2020interplanetary} extends IPFS, using a longest-chain blockchain
and a cryptocurrency (FIL) to incentivize storage nodes.
Publishers acquire storage contracts with a few nodes, and payments are made in the
cryptocurrency. Filecoin mitigates the risk that these nodes delete the replicas by
requiring storage nodes to hold differently encoded copies of the file, and performing
challenges against each other for the encoded files. These copies are encoded in such a
way that it is slow to reproduce them from the original copy, to avoid relay attacks. As
a result, if the user wants to access the original file, they need to wait a long time for
the decoding of a copy, unless some storage node has a hot copy. Since there is no
in-built incentive for storing hot copies, this service usually costs extra.

Arweave~\cite{williams2019arweave} mitigates slow reads through a Proof-of-Access
algorithm that incentivizes storage nodes to store as many files as possible to
maximize rewards. This is implemented in conjunction with a full replication strategy,
and results in replication levels almost equal to those of classic state machine replication.
Additionally, the system only allows files to be stored \enquote{forever}, through a mechanism
of pre-payment---which lacks the flexibility to control lifetime and deletion, and is
capital-inefficient since payment is~upfront.

In contrast to Filecoin and Arweave, \sysname uses erasure coding to maintain a very low
overhead of $4.5\times$ while ensuring that data survives the loss of up to $2/3$ of the
shards and that the system continues to accept writes even if up to $1/3$ of the shards
are unresponsive.
Furthermore, \sysname does not implement its own separate blockchain for node management but uses Sui instead.

Storj~\cite{storj2018storj} is a distributed storage solution that leverages encoding to
achieve a low replication factor. Its deployed network implements a Reed-Solomon-based
erasure-coding scheme with a $29/80$ configuration~\cite{storj-file-repair}, wherein a file is
encoded into 80 parts, with any 29 sufficient for reconstruction. This approach results in a
$2.75\times$ replication factor, offering a substantial reduction in storage costs compared to
prior systems.
However, a key limitation is its inability to efficiently heal lost parts. The system
relies on users to reconstruct the full file and subsequently re-encode it to facilitate
the recovery of lost parts.
In contrast, \sysname's use of \redstuff incorporates an efficient reconstruction
mechanism, which is critical for the healing property of the erasure-coding scheme, especially under the
churn that occurs naturally in a permissionless system. %
\redstuff builds on the Twin-code framework~\cite{rashmi2011enabling,marina2015security}, which uses two
linear encodings of data to enhance the efficiency of sliver recovery. However, unlike
Twin-code, \redstuff encodes data across
differently sized dimensions and integrates authenticated data structures, achieving
Write Completeness (as defined in \cref{def:cds}) and ensuring Byzantine fault tolerance.

Modern blockchains provide some storage, but it is prohibitively expensive to store
larger blobs due to the costs of full replication across all validators, as well as
potentially long retention times to allow verifiability. Within the Ethereum ecosystem
specifically, the current scaling strategy around L2s involves posting blobs
on the main chain that represent bundles of transactions to be executed and verified
via either zero-knowledge or fraud proofs. Specialized networks, such as
Celestia, based on availability sampling~\cite{al2021fraud}, have emerged to fulfill this
need off the main Ethereum chain. In Celestia, two-dimensional Reed-Solomon codes are
used to encode blobs, and symbols are distributed to light nodes to support \enquote{trustless}
availability. However, all blobs are fully replicated across the validators of the
system for a limited time period of about a month. \sysname instead offers proofs of
availability with arbitrarily long retention periods and a reduced cost of storage per
node, which allows the system to scale inexpensively.

The most closely related work to ours is Semi-AVID~\cite{nazirkhanova2022information},
which has also been explored as an alternative to provide data availability for rollups.
It is similar to the Strawman II design, meaning that although it can achieve the critical
property of verifiable data storage, it cannot achieve Write Completeness unless the full
data is reconstructed. This makes it prohibitively expensive for epoch change and only
suitable for either permissioned systems with no churn or short-lived data storage. This
is the main challenge with AVID~\cite{cachin2005asynchronous}, which is optimized to
provide verifiability of data storage, disallowing the output of $\bot$. This is
overkill, as a malicious writer can simply encode garbage data instead of a failed
encoding. Hence, the machinery to detect and reject failed encodings that AVID provides
is unnecessarily constraining and expensive.

\section{Conclusion}

We introduce \sysname, a novel approach to decentralized blob storage that
leverages fast erasure codes and modern blockchain technology. By utilizing the
\redstuff encoding algorithm and the Sui blockchain, \sysname achieves
high resilience and low storage overhead while ensuring efficient data management and
scalability. Our system operates in epochs, with all operations sharded by $\id$, enabling
it to handle large volumes of data. The innovative two-dimensional BFT encoding protocol
of \redstuff allows for efficient data recovery, load balancing, and dynamic availability
of storage nodes, addressing key challenges faced by existing decentralized storage systems.

Furthermore, \sysname provides
a committee reconfiguration protocol that guarantees uninterrupted data availability during
network evolution. %
Our contributions include
defining the problem of Asynchronous Complete Data Storage, presenting the \redstuff protocol,
and providing a full end-to-end system design with \sysname, including a production-grade
implementation and deployment, paving the way for future advancements in decentralized
storage~technologies.

	\section*{Acknowledgments}
This work was funded by Mysten Labs.
We would like to express our gratitude to Dmitri Perelman, He Liu, Pei Deng, Sadhan Sood, William Bradley, and Zhe Wu for their invaluable contributions to bringing \sysname to production %
and to Damir Shamanaev for his assistance in constructing the smart contracts that connect \sysname with the Sui blockchain.
We extend a special thank you to Joachim Neu for identifying a serious vulnerability in our previous testnet implementation. This vulnerability was related to its use of RaptorQ for erasure coding and led us to replace RaptorQ with RS codes.

\bibliographystyle{ACM-Reference-Format}
\bibliography{bibliography}

\balance

\end{document}